\documentclass[mathleft]{an}
\usepackage{graphicx}
\usepackage{times}
\begin{document}

\Pagespan{789}{}
\Yearpublication{2006}%
\Yearsubmission{2005}%
\Month{11}%
\Volume{999}%
\Issue{88}%

\title{Wind braking of magnetars:\\
to understand magnetar's multiwave radiation properties}

\author{H. Tong\inst{1}\fnmsep\thanks{Corresponding author:
  \email{tonghao@xao.ac.cn}\newline}
  \and R. X. Xu\inst{2}
}
\titlerunning{Wind braking of magnetars}
\authorrunning{Tong \& Xu}
\institute{Xinjiang Astronomical Observatory, Chinese Academy of Sciences, Urumqi, Xinjiang, China
\and School of Physics, Peking University, Beijing, China}

\received{30 May 2005}
\accepted{11 Nov 2005}
\publonline{later}


\abstract{%
Magnetars are proposed to be peculiar neutron stars powered by their super strong magnetic field. Observationally, anomalous X-ray pulsars and soft gamma-ray repeaters are believed to be magnetar candidates. While more and more multiwave observations of magnetars are available, unfortunately, we see accumulating failed predictions of the traditional magnetar model. These challenges urge rethinking of magnetar.
Wind braking of magnetars is one of the alternative modelings. The release of magnetic energy may generate a particle outflow (i.e., particle wind),
that results in both an anomalous X-ray luminosity ($L_{\rm x}$) and significantly high spindown rate ($\dot P$).
In this wind braking scenario, {\em only} strong multipole field is necessary for a magnetar (a strong dipole field is no longer needed).
Wind braking of magnetars may help us to understand their multiwave radiation properties,
including (1) Non-detection of magnetars in Fermi-LAT observations,
(2) The timing behaviors of low magnetic field magnetars, (3) The
nature of anti-glitches, (4) The criterion for magnetar's radio emission, etc.
In the wind braking model of magentars, timing events of magnetars should always
be accompanied by radiative events.
It is worth noting that the wind engine should be the central point in the research since other efforts with any reasonable energy mechanism may also reproduce the results.}

\maketitle

\section{Introduction}

Pulsars are rotating magnetized neutron stars. They are the end product of massive stars.
Since the first discovery of pulsars in 1967 (Hewish et al. 1968), more and
more kinds of pulsar-like objects are found. According to their energy sources, pulsars may be cataloged into
four classes.
\begin{enumerate}
 \item Rotation-powered pulsars. These include radio pulsars (including millisecond pulsars),
 rotation-powered X-ray pulsars, and gamma-ray pulsars.
 \item Accretion-powered pulsars. For neutron stars in a binary system, accretion may power
 both their persistent and burst emissions.
 \item Magnetars. Anomalous X-ray pulsars and soft gamma-ray repeaters are thought to be
 neutron stars powered by their super strong magnetic field.
 \item Thermal-powered neutron stars. If neither of the above sources is available, then the neutron
 star can only radiation thermal photons (since it has a non-zero temperature). X-ray dim isolated neutron
 stars are thought to thermal-powered neutron stars.
\end{enumerate}
Different energy sources may be at work in one source, e.g., there can be thermal emission in rotation-powered
pulsars. Magnetar is a special kind of pulsar-like objects. They are discovered by the progress of
multiwave observations of pulsars (X-ray observations rather than the traditional radio observations, Kouviotou et al. 1998).
We are beginning to know more and more of magnetars in recent years.
The study of magnetars may provide one way to unify different kinds of pulsar-like objects (Kaspi 2010).

\subsection{Basics of magnetars}

\begin{figure}[!htbp]
\centering
 \includegraphics[width=0.45\textwidth]{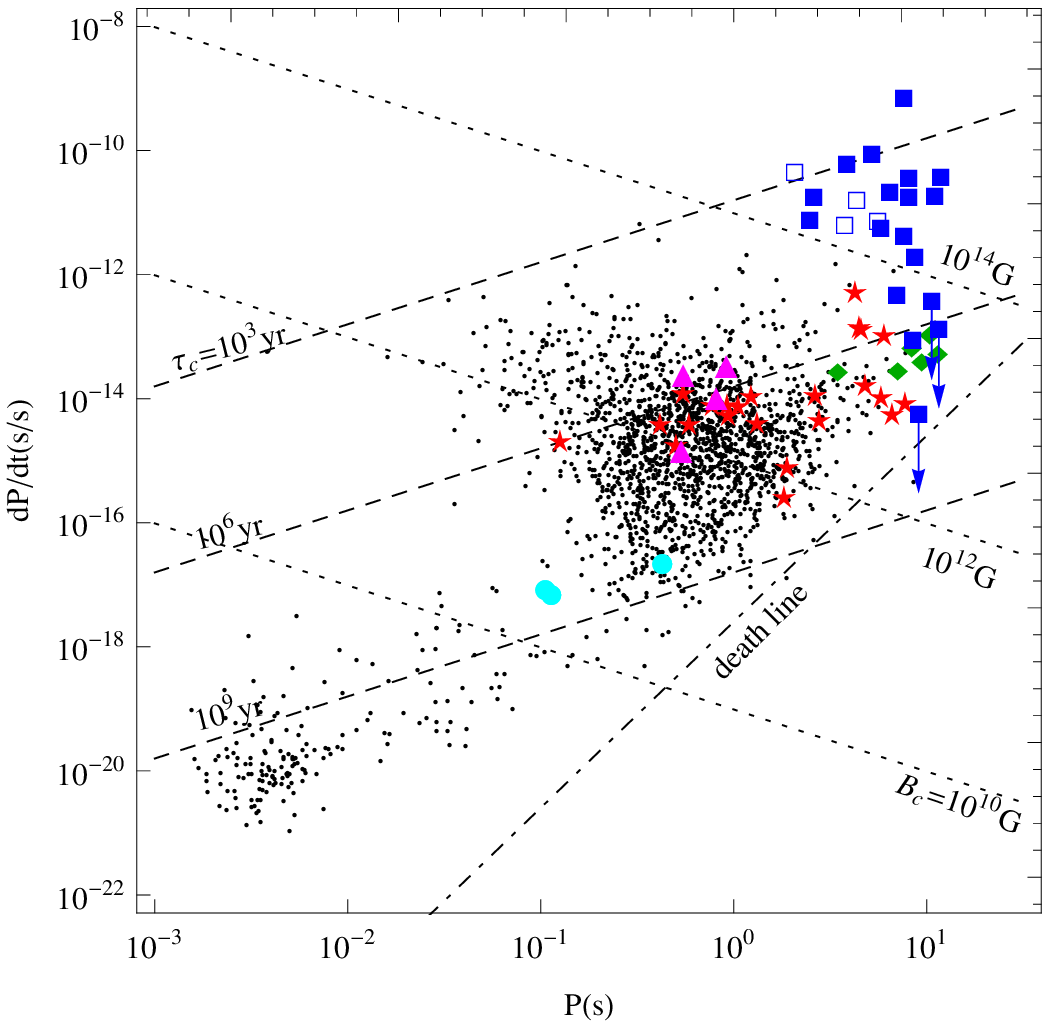}
\caption{Distribution of magnetars on the period period-derivative diagram of pulsars.
Blue squares are magnetars, while empty squares are radio-loud magnetars.
Green diamonds are X-ray dim isolated neutron stars, cyan circles are central
compact objects, red stars are rotating radio transients, magenta triangles
are intermittent pulsars, and black dots are rotation powered pulsars
(including normal pulsars and millisecond pulsars).
Figure 1 in Tong \& Xu (2011), with updates.}
\label{fig_PPdot}
\end{figure}

Anomalous X-ray pulsars (AXPs) and soft gamma-ray repeaters (SGRs) are believed to magnetars. They got
their names due to historical reasons (Mereghetti 2008). Since 1970s, people know that there are two kinds of X-ray pulsars:
rotation-powered X-ray pulsars (e.g., X-ray emissions of Crab and Vela pulsar) and accretion power X-ray pulsars
(accreting neutron stars in binary system).
AXPs have a X-ray luminosity higher (e.g., $10^{35} \,\rm erg\, s^{-1}$) their rotational energy loss rate.
Therefore they can not be rotation-powered.
At the same time, no binary signature is seen in AXPs. Then they are also not accretion-powered.
The energy source of their X-ray emission is unknown at early times. Therefore, they got the name
``anomalous X-ray pulsars''. SGRs are recurrent bursts. Compared with classical gamma ray burst,
SGRs' typical photon energy is lower. Therefore, they are named ``soft gamma-ray repeaters''.
Up to now, we know that AXPs and SGRs belong to the same class of objects. They may be magnetars.
Magnetars form a distinct kind of pulsar-like objects compared with normal pulsars. This can be seen
directly form their distribution on the period period-derivative diagram of pulsars
(Figure \ref{fig_PPdot}).

The first giant flare of magnetars was observed in 1979 (Mazets et al. 1979). The magnetar idea
(neutron stars with magnetic fields as high as $10^{14} -10^{15} \,\rm G$) was proposed by several
authors in 1992 (Duncan \& Thompson 1992; Usov 1992; Paczynski 1992). It was Paczynski (1992) who
pointed that the super-strong magnetic field may explain the super-Eddington luminosity of the 1979
giant flare. Timing observation of the period and period derivative of one SGR (by RXTE) was thought
to be the confirming evidence of magnetars (Kouveliotou et al. 1998). By assuming that the nuetron star is slowed
down by emitting magnetic dipole radiation, the neutron star's surface magnetic field is (Tong et al. 2013a)
\begin{equation}
B=3.2\times 10^{19} \sqrt{P \dot{P}} \, \rm G,
\end{equation}
where $B$ is the star's surface magnetic field, $P$ is the pulsation period,
$\dot{P}$ is the period derivative. For SGR 1806$-$20, its period and period derivative are
$7.47\,\rm s$ and $8.24\times 10^{-11}$, respectively (Kouveliotou et al. 1998). According
to the magnetic dipole braking assumption, SGR 1806$-$20 is a neutron star with age of
1500 years\footnote{The characteristic age is defined as $P/(2\dot{P})$.}
and surface magnetic field as high as $8\times 10^{14} \,\rm G$. Therefore, it is a young neutron
star with super-strong magnetic field\footnote{For normal pulsars, their typical magnetic field
is $\sim 10^{12} \,\rm G$}(i.e., magnetar). Later, not only SGRs but also AXPs are thought to be magnetars.
In 2008, the traditional magnetar model was (Mereghetti 2008): 1) Magnetars are young neutron stars;
2) These neutron stars have dipole magnetic field higher than
the quantum critical field\footnote{The quantum critical field is defined as when the electron cyclotron
energy equals its rest mass energy, $B_{\rm q} =\frac{m_{\rm e}^2 c^3}{e \hbar} =4.4\times 10^{13} \,\rm G$.
In a magnetic field higher than the quantum critical value, quantum electrodynamics must be employed to
treat the microscopic processes.}; 3) The multipole field of these neutron stars may be even higher, e.g.,
as high as $10^{14}-10^{15} \,\rm G$. The dipole field of magnetars provides the braking torque,
while the multipole field is responsible for the burst, super-Eddington luminosity, and persistent emissions
of AXPs and SGRs.

Since 2006, more and more multiwave observations of magentars are available (from radio to optical and IR,
soft X-ray and hard X-ray, and gamma-ray etc). There are observations which are consistent with the magnetar model.
These observations are for the magnetar model if AXPs and SGRs are magnetars.
Therefore, these observations
are model depend evidences for the magnetar model (Tong \& Xu 2011).
Meanwhile, we see accumulating
evidences of failed predictions of the traditional magnetar model (Tong \& Xu 2011).
The discovery of a low magnetic field magnetar in 2010 challenged the traditional magnetar model directly
(Rea et al. 2010; Tong \& Xu 2012). The low magnetic field magnetar (SGR 0418+5729) is an old neutron star
with surface dipole field less than $7.5\times 10^{12} \,\rm G$.
But at the same time, it can have magnetar-like
activities.
It will not be too incorrect to say that
none of the predictions of the traditional magnetar model is observed.
The failed predictions of the traditional magnetar model require rethinking the magnetar idea.
There are 3+1 things to do concerning magnetars
\begin{enumerate}
  \item What's the origin of strong magnetic field in magnetars and pulsars? This is relevant to whether AXPs and SGRs
  are magnetars or not.
  \item What's the emission mechanism of magnetar multiwave radiation properties? The multiwave emission mechanism of magnetars remains illusive.
  \item The birth and environment of magnetars. The environment of magnetars possibly includes: fallback disks,
  pulsar wind nebulae, supernova remnants, and binary companions (if the magnetar is in a binary system).
  \item The relation between magnetars and other pulsar-like objects. Magnetars are just a special kind of pulsars. Therefore, we must understand various pulsar-like objects at the same time. We want to know what's the relation between magnetars and X-ray dim isolated neutron stars (XDINSs), central compact objects (CCOs), high magnetic field pulsars (HBPSRs), and most importantly normal pulsars.
\end{enumerate}

There exist several alternative modelings of magnetars (Tong \& Xu 2011).
\begin{enumerate}
  \item Twisted magnetosphere model (Thompson, Lyutikov \& Kulkarni 2002). The magnetar magnetosphere may be globally twisted.
  And a partially twisted magnetosphere model was also investigated (i.e., corona model of magnetars, Beloborodov \& Thompson 2007; Beleborodov 2009).
  \item Wind braking of magnetars (Tong \& Xu 2012; Tong et al. 2013a). In the wind braking model, magnetars are neutron stars with strong multipole field. A strong dipole field is no longer needed. A particle outflow dominates the rotational energy loss rate of magnetars. The multipole field is responsible for the braking torque, persistent and burst emissions of magnetars. In the wind braking model of magnetars,
      timing events of magnetars should always be accompanied by radiative events.
  \item Magnetothermal evolution model (Vigano et al. 2013). Coupled evolution of magnetic field and temperature
  of neutron stars may explain the surface thermal emission of various kinds of pulsar-like objects.
  \item Fallback disk model (Alpar 2001; Alpar, Ertan \& Kaliskan 2011). A neutron star with a fallback disk may explain some aspects of magnetars. And there is already a disk found in AXP 4U 0142+61 (through optical/IR observations, Wang, Chakrabarty \& Kaplan 2006).
  \item Accretion induced star-quake model (Xu, Tao \& Yang 2006; Xu 2007).
  The self-confined quark star surface can explain the super-Eddington luminosity of
  magnetar giant flares. Accretion from a fallback disk
      is responsible for the spindown and persistent emissions. A quark star with a fallback disk may provide another way
      unifying different kinds of pulsar-like objects.
  \item Quark-nova remnant model (Ouyed, Leahy \& Niebergal 2007, 2011).
  After the supernova, there may be a transition from a
  neutron star to a quark star. This is dubbed as a quark-nova.
  A quark star with some kind of quark-nova remnant may explain
  several pulsar-like objects (including AXPs and SGRs).
  \item White dwarf model (Paczynski 1990; Malheiro, Reuda \& Ruffini 2012).
  If the central star of AXP and SGR is a white dwarf,
  then the rotational energy of the white dwarf is enough to power the persistent emissions of AXPs and SGRs.
\end{enumerate}
The first three models are in the magnetar domain (i.e., they all involve neutron stars powered by strong magnetic field).
The last four models are more or less beyond the magnetar model.

\section{Toward an understanding of magnetar multiwave radiation properties}

\subsection{Non-detection in Fermi-LAT observations}
In the traditional model of magnetars, magnetars are neutron stars with both strong dipole field and strong multipole
field. Although the magnetic field at the magnetar surface is very high, the magnetic field in the outer
magnetosphere is relatively low. Therefore, particles may be accelerated in the outer magnetosphere
and magnetars are expected to be high-energy gamma-ray emitters detectable by Fermi-LAT
(Cheng \& Zhang 2001). However, the X-ray luminous AXP 4U 0142+61 is not detected in Fermi-LAT observations
(Sasmaz Mus \& Gogus 2010). There is also no significant detection in Fermi-LAT observations of all
AXPs and SGRs (Abdo et al. 2010). Then there are conflicts between outer gap model in the case of magnetars
and Fermi-LAT observations (Tong, Song \& Xu 2010, 2011).
AXP 4U 0142+62 should have been detected by Fermi-LAT. The present observational
upper limits are already below the theoretical calculations for some parameter space.
There are possibly two solutions for this conflict:
\begin{enumerate}
  \item AXPs and SGRs are fallback disk systems. Then most of them are not expected to be 
  gamma-ray emitters.
  \item AXPs/SGRs are magnetars braked down by a particle wind. If a particle outflow dominates the magnetar's
  rotational energy loss rate, then the corresponding surface dipole field can be much lower (i.e., 10-100 times smaller).
  Meanwhile, in the presence of a particle wind, vacuum gaps can not exist in the magnetosphere.
\end{enumerate}
Fermi deeper observations may help us to distinguish between the fallback disk model and magnetar model
for AXPs and SGRs.

\subsection{Hard X-ray emission cutoff}

The soft X-ray spectral of magnetars are uausally made up of two components: a blackbody component (with temperature
$\sim 0.5 \,\rm keV$) and power law component (with a photon index $\Gamma \sim 3-4$). Extrapolating the soft
X-ray components, magnetars are not expected to luminous in the hard X-ray range. However,
INTEGRAL observations found that many magnetars are detected in hard X-ray (Gotz et al. 2006).
The hard X-ray can be fitted with a power law with photon index $\Gamma \sim 1$.
 And the hard X-ray energy output is about half the magnetar's total
electromagnetic energy output.
Therefore, the hard X-ray component of magnetars is a distinct component compared
with the soft X-ray component. And it is an indispensable part of the magnetar's energy budget.
There are various proposals for the origin of hard X-ray emission both in the magnetar
model and the fallback disk model.
And the hard X-ray emission cutoff is crucial to distinguish between different models.

A possible cutoff in the hard X-ray emission of AXP 4U 0142+61
is reported recently (Wang, Tong \& Guo 2013).
Using nearly nine years INTEGRAL observations, a possible cutoff of $\approx 130 \,\rm keV$ is seen.
With a cutoff of $130\,\rm keV$, we can rule out hard X-ray emission models involving ultra-relativistic
electrons. Both the microscope and bulk motion of electrons should be at most mildly relativistic.
During the nine years interval, the total hard X-ray luminosity is relatively stable.
Therefore, a persistent source of electrons are needed rather than transient.
Therefore, there must exist a persistent component of particle outflow.
Hard X-ray Modulation Telescope (known HXMT, by China) can determine the cutoff energy more accurately
in the future.

\subsection{Soft X-ray timing behavior}

\subsubsection{Wind braking of magnetars}

In timing study of magnetars, the magnetic dipole braking assumption is often employed (Kouveliotou et al. 1998).
However, the magnetic dipole braking assumes an perpendicular rotator in vacuum.
Therefore, it is just an pedagogical model (Li et al. 2013). The non-detection of magnetars by Fermi-LAT,
the timing difference between magnetars and high magnetic field pulsars, and most importantly the
varying period derivative of magnetars, these observations may imply that magnetars have a different
braking mechanism from that of normal pulsars (Tong et al. 2013a).
Both pulsars and magnetars should be braked down by a
particle wind (i.e., a mixture of particles and electromagnetic fields). The difference between them
is that for pulsars magnetic dipole braking is valid to the lower order approximation.
The particle wind mainly causes higher order timing effects, e.g., braking index, timing noise.
However, for magnetars magnetic dipole braking is incorrect even to the lowest order approximation.
Therefore, in timing study of magnetars we must employ the full formalism of wind braking.

The soft X-ray luminosity of magnetars $L_{\rm x}$ originates from their magnetic field decay.
During the decay of magnetic field, a particle outflow may also be generated\footnote{There
must exist some amount of nonthermal particles because magnetars have nonthermal emissions, e.g.,
radio, optical, nonthermal soft X-ray and hard X-ray etc.} (i.e., particle wind).
A natural estimation of particle wind luminosity is that $L_{\rm p} = L_{\rm x}$ (the particle
wind luminosity equals the soft X-ray luminosity). When the particle wind luminosity is known,
we can calculate the spindown behaviors of magnetars in the wind braking scenario.
The dipole magnetic field of magnetars in the case of wind braking is (Tong et al. 2013a)
\begin{eqnarray}\label{Bdipw}
\nonumber
 B &=& 4.0\times 10^{25} \frac{\dot{P}}{P} \,L_{\rm p,35}^{-1/2} \,{\rm G}\\
&=& 4.0\times 10^{13} \frac{\dot{P}/10^{-11}}{P/10\,\rm{s}} \,L_{\rm p,35}^{-1/2} \,{\rm G},
\end{eqnarray}
where $L_{\rm p,35}$ is the particle wind luminosity in units of $10^{35} \,\rm erg s^{-1}$.
In the wind braking scenario of magnetars, a strong dipole field is no longer needed.
Magnetars are neutron stars with strong multipole field. The particle wind luminosity may
have significant variations (as that of their X-ray luminosities). This may explain why
many magnetars have a varying period derivative and other timing events.
Since both the soft X-ray luminosity and the particle wind are from the
magnetic field decay, the timing events of magnetars should always be accompanied
by radiative events in the wind braking model.

\subsubsection{Timing behaviors of low magnetic field magnetars}

The discovery of low magnetic field magnetar SGR 0418+5729 has challenged the traditional magnetar model directly
(Rea et al. 2010). A low magnetic field magnetar is thought to be a neutron star with
relatively low surface dipole field (e.g., $\sim 10^{12} \, \rm G$, in order to explain the timing behavior)
and much higher multipole field (e.g., $>10^{14} \,\rm G$, in order to explain the persistent and burst emisions).
However, when calculating the surface dipole field, the magnetic dipole braking assumption is employed.
The real case may include both a dipole radiation component and a particle wind component.
For SGR 0418+5729, its particle wind component may have been ceased. The magnetic field
which is responsible for the star's spinning down is effectively $B\sin\theta$, where $\theta$
is the angle between the magnetic axis and the rotation axis (i.e., magnetic inclination angle).
If SGR 0418+5729 has a small inclination angle, e.g., $\theta=5^{\circ}$, its surface dipole
may be as high as $10^{14}\,\rm G$. Therefore, SGR 0418+5729 may be a normal magnetar instead of
a low magnetic field magentar (Tong \& Xu 2012).
It has a small period derivative because its magnetic inclination angle is small.

The second low magnetic field magnetar Swift J1822.3$-$1606 has different period derivatives reported
(Rea et al. 2012a; Scholz et al. 2012). In the wind braking model of magnetars, the particle wind luminosity
decreases with time after the outburst. This may result in a decreasing period derivative. Therefore,
different period derivatives are obtained using different time span of timing observations
(Tong \& Xu 2013). Meanwhile, the
fluctuation of the particle wind is also responsible for the large timing noise.
Subsequent timing study
can tell us whether wind braking is important in this source or not.

\subsubsection{Anti-glitch of magnetars}

Pulsar are very stable clocks in the universe. At same time, detailed studies found several timing
irregularities in pulsars: glitch (sudden spin-up of the pulsar) and timing noise etc.
Up to now hundreds of glitches are
observed in hundreds of pulsars (including several magnetars).
All these glitches are spin-up events. Recently, an anti-glitch
is reported in one magnetar (a spin-down event, Archibald et al. 2013). If confirmed by future observations,
anti-glitches may require rethinking of glitch modeling of all neutron stars. Observationally,
the anti-glitch is accompanied by an outburst event. The particle wind luminosity is higher during
the outburst than during the persistent state. A stronger particle particle wind will cause
a higher spindown rate during the outburst. After some time, a net spindown of magnetar is expected
(i.e., anti-glitch). Therefore in the wind braking scenario, there are no anti-glitches.
Anti-glitch is just a period of enhanced spindown. If there are enough timing observations,
a period of enhanced spindown rate is expected (Tong 2013). A second anti-glitch event
will help us to discriminate between different models.

Considering that the anti-glitch may be caused by an enhanced particle wind, the opposite
case is also possible. During a time interval, the star's particle wind may be lower (or even
ceased). After sometime, the star will look like to have a net spin-up. Observationally,
this corresponds to the timing behavior of intermittent pulsars.
The spindown behavior of intermittent pulsars is
understandable in the pulsar wind model (Li et al. 2013). Therefore,
both anti-glitch and the spin-down behavior of intermittent pulsars can be
understood uniformly in the wind braking scenario.

\subsection{Criterion for magnetar's radio emission}

Originally, magnetars are expected to be radio quiet both in the magnetar model and the
fallback disk model. However, transient pulsed radio emission from one magnetar was
discovered in 2006 (Camilo et al. 2006). There are distinct properties of magnetar
radio emissions: 1) Their flux and pulse profile vary with time. 2) They have a flat
spectrum in the radio band. 3) The radio emission is transient in nature (with duration
of years). We even do not know whether the magnetar's radio emission is from their magnetic
energy or rotational energy.
With three radio emitting magnetars at hand, the empirical
``fundamental plane of magnetar radio emission'' was proposed (Rea et al. 2012b).
Rea et al. (2012b) proposed that a magnetar is radio-loud if and only if its persistent
X-ray luminosity is smaller than its rotational energy loss rate.
And the magnetar radio emission should
come from their rotational energy. However, this proposal failed in one new source
Swift J1834.9$-$0846 (Tong, Yuan \& Liu 2013b). Swift J1834.9$-$0846 has persistent X-ray
luminosity smaller than its rotational energy loss rate. Therefore, it should have radio emissions
if the fundamental plane of magnetar radio emission is correct. However, it is not
detected in radio using Nanshan 25 meter radio telescope
(of Xinjiang Astronomical Observatory, Chinese Academy of Sciences).
Green Bank Telescope also
reported non-detection of this source (see references in Tong et al. 2013b). 
We observed this source using GMRT in 2013 January,
which is also not detected. Therefore, at present we can only say that ``low luminosity magnetars
are more likely to have radio emissions'' (Tong et al. 2013b).
And the magnetar radio emission should come from their
magnetic energy.

The reason why low luminosity magnetars are more likely to have radio emissions
may be that they are more like to have similar magnetosphere to that of normal
radio pulsars. According to the wind braking model of magnetars (Tong et al. 2013a),
for low luminosity magnetars, the magnetic dipole braking assumption is correct to
the lowest order approximation (the same as that of normal radio pulsars).
Therefore, during the persistent state, a magnetosphere similar to that of normal radio
pulsars is prepared. Then it is natural that they may have radio emissions.

\section{Summary and prospect}

Multiwave observations, especially high-energy observations, have discovered increasing
kinds of pulsar-like objects. Among them, magnetars form a different population from
that of radio pulsars. The magnetar model may provide one way to understand different
kinds pulsar-like objects. Originally, magnetars are thought to be neutron stars with
superstrong dipole field. This must be wrong since there exist high magnetic field
pulsars. Later, magnetars are thought to be neutron stars with both strong dipole field
and strong multipole field. However, this is also incorrect because we have discovered
several low magnetic field magnetars. We now know that the key difference between
normal pulsars and magnetars is the absence or presence of strong multipole field.
Normal pulsars are neutron stars without strong multipole field, while magnetars
are neutron stars with strong multipole field. A strong dipole field is no longer
needed in the wind braking model of magnetars. The decay of multipole field
will generate a particle outflow (i.e., particle wind). This particle wind is responsible
for both the spindown and multiwave radiation (at least nonthermal radiations) of magnetars.
The wind braking model of magnetars (Tong et al. 2013a) may explain the correlation
between magnetar timing and radiation properties, e.g., decreasing period derivative
after outburst, period derivative variations during the persistent state (radiation
flux variations are also observed). In general, in the wind braking model, the timing
events of magnetars should always be accompanied by radiative events. The timing
of low magnetic field magnetars, anti-glitch of magnetars, can be understood safely
in the wind braking model. The existence and property of particle wind can also
help to explain the high-energy gamma-ray, hard X-ray, and radio observations
of magnetars. More investigations of the wind braking model of magnetars
and more multiwave observations can tell us whether magnetars are wind braking
or not.

The wind braking model and other alternatives (e.g., the corona model of magnetars)
share some merits. One point is that: once the particle outflow is generated, the
subsequent plasma process does not depend on where the particle comes from, i.e., the wind engine.
Besides that the particles may be generated by magnetic field decay,
other energy mechanisms (e.g., star quake) could also be the source of power.
At present, we discuss the following two possibilities for the source of particle wind.
\begin{enumerate}
  \item The magnetar model. The magnetic energy is the ultimate energy source.
  Magnetic activities (e.g., trigger by seismic activities) is responsible for
  the star's activities. In the magnetar model, the free parameters are:
  the dipole field and multipole field (or poloidal and toroidal field) in the magnetosphere,
  the dipole field and multipole field in the crust, and the crust shear modulus etc.
    \item The quark star with fallback disk model. Quark star may be more stable than
    neutron star. The self-bound quark star surface can explain the super-Eddington
    luminosity of magnetars naturally. A quark star with fallback disk can explain
    the persistent radiation and timing, and burst properties of AXPs and SGRs.
    Meanwhile, fallback disk systems may also unify different kinds of pulsar-like
    objects. The free parameters in the quark star with fallback disk model are:
    surface dipole field, accretion rate, the shear modulus,
    and quark star mass, etc. In this regime, the puzzling behaviors of anomalous X-ray pulsars and soft gamma-ray repeaters relate essentially to the challenging problem: the equation of state of cold matter at supra-nuclear density (Xu 2011).
\end{enumerate}
The underlying energy source is the fundamental problem in studying pulsar-like
objects. Whether magnetars exist or not, whether quark stars exist or not,
they are both big problems in pulsar researches.

\subsection{Prospect: magnetars in astrophysics}

From Figure \ref{fig_PPdot} we see that there are various kinds of pulsar-like objects.
The study of magnetars may help us to achieve the ``grand unification of neutron stars''
(Kaspi 2010). At present, we are far from the ultimate truth. If we try to understand
pulsar-like objects from the magnetar point of view,
we may have the following picture of magnetars in astrophysics:
\begin{enumerate}
\item
The currently observed anomalous X-ray pulsars and soft gamma-ray repeaters are magnetars (neutron stars with strong multipole field).
\item
X-ray dim isolated neutron stars may be dead magnetars. Their simple thermal
X-ray spectrum may require a magnetized neutron star atmosphere.
\item
The central compact objects will become magnetars in the future or
they are disk braked magnears. The X-ray hot spot may due to the
presence of strong crustal field or due to accretion heated polar cap.
\item
High magnetic field pulsars can also have magnetar-like activities, e.g., bursts (as has been
already observed). Since they have high surface dipole field,
some of them may also have strong multipole field.
\item
A normal pulsar can also have magnetar-like activities although with a low probability.
Considering that the total number of normal pulsars is significantly larger than magnetars, therefore
observing one burst (or outburst) from normal pulsars is not impossible.
\item
If a magnetar is born in a binary system, we may see an accreting magnetar.
The key point is can we see magnetism-powered activities in accreting systems?
An accreting neutron star with high surface dipole field should not be called
an accreting magnetar. It is just an accreting high magnetic field pulsar.
\item
The formation of magnetars in other galaxies may be one of the central engine of gamma-ray burst.
The large rotational
energy of the rapidly rotating magnetar is the energy source. The magnetar strength
magnetic field can extract the star's rotational energy in a short time scale.
And this sudden energy release will result in a gamma-ray burst
\end{enumerate}

\acknowledgements
Figure 1 is reused by permission of:
``Magnetars: fact or fiction?, Tong \& Xu, IJMPE, Vol. 20, Supplement 2, 15-24, Copyright@2011,
World Scientific Publishing Company''.
H.Tong is supported by NSFC (11103021), West Light Foundation of CAS (LHXZ201201), Xinjiang Bairen project, and Qing Cu Hui of CAS. R.X.Xu is supported by National Basic Research Program of China (2012CB821800), NSFC (11225314), and XTP project XDA04060604.


\end{document}